\documentclass[12pt,a4paper]{article}
\usepackage{amsfonts}
\usepackage{amssymb}
\usepackage{latexsym}
\usepackage{graphicx}
\begin{document}

\begin{center}
{\Large \bf Single-brane world\\ with stabilized extra dimension} \\

\vspace{4mm}

Mikhail N. Smolyakov, Igor P.~Volobuev\\ \vspace{0.5cm} Skobeltsyn Institute of
Nuclear Physics, Moscow State University
\\ 119991
Moscow, Russia\\
\end{center}

\begin{abstract}
We present a model describing a single brane with tension embedded
into a five-dimensional space-time with compact extra dimension,
which can be easily stabilized. We  examine the linearized gravity
in the model and obtain an expression for the four-dimensional
Planck mass on the brane in terms of the model parameters. It is
also shown that the scalar sector of the effective
four-dimensional theory contain a tachyonic mode, and we discuss
the problem of stability of the model.
\end{abstract}

\section{Introduction}
Brane world models and their phenomenology have been widely
discussed in the last years. One of the most interesting brane
world models is the Randall-Sundrum model with two branes, -- the
RS1 model \cite{Randall:1999ee}. This model solves the hierarchy
problem due to the warp factor in the metric and predicts an
interesting new physics in the $TeV$ range of energies.

Most of the brane world models with one compact extra dimension
and thin branes with tension demand the existence of at least two
branes. At the same time the matter located on the brane, which is
not "our"\ brane, can strongly affect the world located on "our"\
brane. For the case of the RS1 model it was shown in
\cite{SV-RS}. So it would be quite interesting to find
out, whether it is possible to construct a model with only one
tensionful brane in a compact extra dimension, admitting a
solution to the hierarchy problem in the way analogous to that
proposed in \cite{Randall:1999ee}.

A characteristic feature of models with single brane is the
presence of at least one tachyonic mode in the perturbative
linearized theory \cite{Lesgourgues:2003mi}. At the same time the
linearized theory, as well as the five-dimensional effective
action describing a brane world model, is valid for the energy
range of the order of the fundamental energy scale of the theory,
defined by the five-dimensional gravity (we suppose that this
scale is of the order of $1-10\,\, TeV$). Thus, if the masses of
the tachyonic modes are far beyond the energy range of its
applicability, their influence on the theory cannot be accessed in
the linear approximation, and one needs to consider the nonlinear
effects.

Some solutions with single brane in a compact extra dimension,
interesting from the cosmological point of view, were obtained in
\cite{Kanti:1999nz}. But  the energy-momentum tensors used
for obtaining these solution are "phenomenological", i.e. they are
added to the action "by hand".

Here we present a model describing the  scalar field minimally
coupled to gravity in a five-dimensional space-time, admitting the
existence of a single brane quite naturally and being of interest
from the point of view of the hierarchy problem. Moreover, the
size of the extra dimension in this model can be easily
stabilized. Thus, the model appears to be devoid of the main flaw
of the original Randall-Sundrum model -- the existence of the
massless scalar mode, called the radion, which arises due to the
fluctuations of the branes with respect to each other and whose
interactions contradict the existing experimental data. We argue
that the stabilization of the size of extra dimension is made in
the same way as in \cite{wolfe}. This method is free from
the main disadvantage of the approach proposed in
\cite{wise}, where the backreaction of the scalar field on
the background metric is not taken into account. There is only one
tachyonic mode in the model with the  mass of the order of the
four-dimensional Planck mass $\sim 10^{19} GeV$, thus lying far
beyond the applicability range of the theory.

The paper is organized as follows. In Section~2 we present the
background solution and the method of its stabilization. In
Section~3 we obtain gauge conditions and equations of motion for
linearized gravity in the model. In Section~4 we consider the
tensor modes and obtain an expression for the four-dimensional
Planck mass on the brane in terms of the model parameters. In
Section~5 we consider the scalar sector of the theory, discuss the
stability of the model and obtain the estimates for the mass of
the lowest scalar mode and its coupling constant to matter on the
brane. And finally, we discuss the obtained results.

\section{The model}
Let us denote the coordinates  in five-dimensional space-time
$E=M_4\times S^{1}$  by $\{ x^N\} \equiv \{x^{\mu},y\}$, $N=
0,1,2,3,4, \, \mu=0,1,2,3 $, the coordinate $x^4 \equiv y, -L\leq
y \leq L$ parametrizing the fifth dimension with identified points
$-L$ and $L$. The brane is located at the point $y=L$.

The  action of stabilized brane world model can be written as
\begin{eqnarray}\label{actionDW}
S=\int d^{4}x \int_{-L}^L dy \sqrt{-g} \left[  2 M^3R -\frac{1}{2}
g^{MN}\partial_M\phi\partial_N\phi-V(\phi)\right]
-\int_{y=L}d^{4}x\sqrt{-\tilde g}\lambda(\phi),
\end{eqnarray}
Here $V(\phi)$ is a bulk scalar field potential and
$\lambda(\phi)$ is the brane scalar field potential,
$\tilde{g}=det\tilde g_{\mu\nu}$, and $\tilde g_{\mu\nu}$  denotes
the metric induced on the brane. The signature of the metric
$g_{MN}$ is chosen to be $(-,+,+,+,+)$.

The standard ansatz  for the  metric and the scalar field, which
preserves the Poincar\'e invariance in any four-dimensional
subspace $y=const$, looks like
\begin{eqnarray}\label{metricDW}
&ds^2=  e^{-2A(y)}\eta_{\mu\nu}  {dx^\mu  dx^\nu} +  dy^2 \equiv
\gamma_{MN}(y)dx^M dx^N,& \\ &\phi(x,  y) = \phi(y),&
\end{eqnarray}
 $\eta_{\mu\nu}$ denoting the flat Minkowski metric. If one substitutes
 this ansatz into the equations corresponding to action
(\ref{actionDW}), one gets a rather complicated system of
nonlinear differential equations for functions $A(y),\phi(y)$:
\begin{eqnarray}\nonumber
\frac{d V}{d\phi}+\frac{d\lambda}{d\phi}\delta(y-L)= -4A'\phi'+\phi'',\\
\nonumber 12M^3 (A')^2+\frac{1}{2}(V-\frac{1}{2} (\phi')^2)=0, \\
\label{yd}
\frac{1}{2}\left(\frac{1}{2}(\phi')^2+V+\lambda\delta(y-L)
\right)=-2M^3\left(-3A''+6(A')^2\right).
 \end{eqnarray}
Here and below $'\equiv{d}/{d{y}}$. An interesting conclusion
following from these equations is that the relation
\begin{equation}
A''(y)=\frac{1}{12M^{3}}{\phi'}^{2}
\end{equation}
 holds in the bulk for any potential $V(\phi)$, and thus
$A''\ge 0$ in the bulk. This inequality was also obtained in
\cite{Freedman} from the weaker energy condition.

To find an analytic solution to this system we will use the
results of \cite{wolfe,Brandhuber}. Let us
consider a special class of potentials, which can be represented
as $$ V(\phi)=\frac{1}{8} \left(\frac{d W}{d\phi}\right)^2-
\frac{1}{24M^3}W^2(\phi). $$ Let us also suppose that
\begin{equation}
W(\phi)=\frac{8\gamma}{3}\phi^{\frac{3}{2}}.
\end{equation}
In this case the scalar field potential takes a simple polynomial
form
\begin{equation}
V(\phi)=\gamma^{2}\left(2\phi-\left(\frac{2}{3M}\right)^{3}\phi^{3}\right),
\end{equation}
and the corresponding continuous background solution can be easily
found (with the help of the procedure described in
\cite{wolfe,Brandhuber}):
\begin{eqnarray}\label{bckgrsol}
\phi=\left(\gamma y\right)^{2},\\
\nonumber A=\frac{1}{36M^{3}}\left(\left(\gamma
y\right)^{4}-\left(\gamma L\right)^{4}\right).
 \end{eqnarray}
The additive constant in the solution for $A(y)$ is chosen in such
a way that the coordinates $\{x^\mu\}$ are Galilean on the brane
(see \cite{Rubakov:2001kp,Boos:2002ik} for
details). We will refer all the energy parameters, which  appear
in the theory, to this Galilean coordinate system on the brane.

In order the equations of motion be valid on the brane too, one
needs to finetune the brane potential $\lambda(\phi)$. We choose
\begin{equation}
\lambda(\phi)=-W(\phi)=-\frac{8\gamma}{3}\phi^{\frac{3}{2}}.
\end{equation}
In this case the brane appears to be of the BPS type. The size of
the extra dimension is not defined by the solution yet.

To stabilize the size of the extra dimension, let us add the
following term to the scalar field potential on the brane:
\begin{equation}
\Delta\lambda(\phi)=\beta^{2}\left(\phi-\phi_{0}\right)^{2}.
\end{equation}
Such an addition will not affect the equations of motion provided
\begin{equation}
\phi|_{y=L}=\phi_{0},
\end{equation}
which means that
\begin{equation}
L=\frac{\sqrt{\phi_{0}}}{\gamma}.
\end{equation}
Thus, we see that the size of the extra dimension is stabilized.

It is necessary to note that the background solution presented
above was obtained without imposing $Z_{2}$ orbifold symmetry,
which is inherent to the most brane world models, although the
solution itself possesses reflection symmetry with respect to the
point $y=0$.

We also suppose that the parameters of the potentials  $\gamma,
\phi_{0}, \beta$, when made dimensionless by the fundamental
five-dimensional energy scale of the theory $M$, should be
positive quantities of the order $O(1)$, i.e. there should be no
hierarchical difference in the parameters. We note that action
(\ref{actionDW}) and the corresponding four-dimensional effective
theory can be used only at the energy scales $E\lesssim M$
measured in Galilean coordinates on the brane.

\section{Linearized gravity}
Now let us turn to the examination of linearized gravity in the
model. We represent the metric and the scalar field as
\begin{eqnarray}\label{metricparDW}
g_{MN}(x,y)&=& \gamma_{MN}(y) + \frac{1}{\sqrt{2M^3}} h_{MN}(x,y),
\\ \label{metricparDW1}
\phi(x,y) &=& \phi(y) + \frac{1}{\sqrt{2M^3}} f(x,y).
\end{eqnarray}
To simplify the analysis, let us impose $Z_{2}$ orbifold symmetry
conditions (although this symmetry is not necessary for obtaining
the background solution). Correspondingly, the metric $g_{MN}$ and
the scalar field $\phi$ satisfy the orbifold symmetry conditions
\begin{eqnarray}
\label{orbifoldsym}\nonumber
 g_{\mu \nu}(x,- y)=  g_{\mu \nu}(x,  y), \quad
  g_{\mu 4}(x,- y)= - g_{\mu 4}(x,  y), \\
   g_{44}(x,- y)=  g_{44}(x,  y), \quad
   \phi(x,- y)=  \phi(x,  y).
\end{eqnarray}

We realize that imposing $Z_{2}$ orbifold symmetry is a rather
artificial procedure. But a consistent and thorough analysis of
linearized gravity without this symmetry, i.e. taking into all the
degrees of freedom coming from the metric,  is a very complicated
problem (for example, we cannot impose the gauge conditions which
will be used later). At the same time, a theory with the orbifold
symmetry makes sense and was studied, for example, in
\cite{Lesgourgues:2003mi}. Moreover, we have a developed
formalism for studying linearized gravity in brane world models
stabilized by the bulk scalar field and with extra dimension
forming the orbifold $S^1/Z_2$ -- see \cite{BMSV1}. The
only difference from this case is  that all the fields should have
a "good"\ behavior at the point $y=0$, i.e. the fields should be
smooth at $y=0$, which corresponds to the absence of the brane as
a physical object at this point. For these reasons, in this paper
we restrict ourselves to the case with $Z_{2}$ orbifold symmetry
conditions.

Substituting representation (\ref{metricparDW}) and
(\ref{metricparDW1}) into action (\ref{actionDW}) and keeping the
terms of the second order  in $h_{MN}$ and $f$, we get the second
variation Lagrangian of this action \cite{BMSV1}. This Lagrangian
is invariant under the gauge transformations
\begin{eqnarray}\nonumber
&h_{MN}^{(\prime)}(x,y)=h_{MN}(x,y)-(\nabla_M{\xi_N}+\nabla_N{
\xi_M}),&\\ \nonumber &f^{(\prime)}(x,y)=f(x,y)-\phi'\xi_4,&
\end{eqnarray}
where $\nabla_M$ is the covariant derivative with respect to the
background metric, provided ${\xi_M(x,y)}$ satisfy the orbifold
symmetry conditions $$ \xi_{\mu}(x,-y)= \xi_{\mu}(x,y), \quad
\xi_4(x,-y)= -\xi_4(x,y). $$ These gauge transformations are a
generalization of the gauge transformations in the unstabilized
RS1 model \cite{Rubakov:2001kp,Boos:2002ik}. We will use them to
isolate the physical degrees of freedom of the fields $h_{MN}$ and
$f$. We also note that since $\xi_{4}|_{y=L}=0$, the brane appears
to be straight (the disadvantages of bent-brane formalism were
discussed in \cite{Arefeva}).

It was shown in \cite{BMSV1} that with the help of these
gauge transformations one can impose the gauge
\begin{eqnarray}\label{gauge}
(e^{-2A}h_{44})'-\frac{1}{3 M^3}e^{-2A}\phi'f=0, \\
\nonumber h_{\mu 4} =0,
\end{eqnarray}
after which there remain the gauge transformations satisfying
\begin{equation}\label{restr}
({e^{2A}\xi_\mu})'=0.
\end{equation}

A substitution
\begin{eqnarray}\label{subst}
h_{\mu\nu} = b_{\mu\nu} - \frac{1}{2} \gamma_{\mu\nu} h_{44}
\end{eqnarray}
allows us to decouple the equations of motion, following from the
second variation Lagrangian, in  gauge (\ref{gauge}). Gauge
transformations satisfying (\ref{restr}) allow one to impose the
traceless-transverse gauge condition on the field
$b_{\mu\nu}$ \cite{Boos:2002ik,BMSV1}
\begin{eqnarray}\label{compl_gauge} \tilde b = \gamma^{\mu\nu}{
b_{\mu\nu}}=0, \quad
\partial^\nu{ b_{\mu\nu}}=0,
\end{eqnarray}
 the residual gauge transformations now being
\begin{equation}\label{ok1}
\xi_\mu=e^{-2A}\epsilon_\mu(x),\qquad
\partial^\nu\epsilon_\nu(x)=0,\qquad \Box{\epsilon_\nu}=0,
\end{equation}
where $\Box=\eta^{\mu\nu}\partial_{\mu}\partial_{\nu}$.
Transformations (\ref{ok1}) act only on the massless mode of the
field $b_{\mu\nu}$ and provide the correct number of degrees of
freedom of the massless graviton \cite{Boos:2002ik}.

Finally, we get the equations of motion in the interval $(0,L)$
with corresponding boundary conditions at the points $y=0$, $y=L$
for the field $b_{\mu\nu}$
\begin{equation}\label{ub}
\frac{1}{2}\left(e^{2A(y)}\Box
{b_{\mu\nu}}+\frac{\partial^2{b_{\mu\nu}}}{\partial
y^2}\right)-b_{\mu\nu}\left(2(A')^2-A''\right)=0,
\end{equation}
\begin{eqnarray}
b'_{\mu\nu}|_{y=+0}=0,\\
\nonumber b'_{\mu\nu}+2A'b_{\mu\nu}|_{y=L-0}=0,
\end{eqnarray}
and for the field $g =e^{-2A(y)}h_{44}(x,y)$
\begin{equation}\label{ug0}
g'' +2g'\left(A'-\frac{\phi''}{\phi'}\right)-\frac{
(\phi')^2}{6M^3} g+\partial_\mu \partial^\mu g =0,
\end{equation}
\begin{eqnarray}\nonumber
g'|_{y=+0}=0, \\ \label{bc} \beta^{2}g' -\partial_\mu \partial^\mu
g|_{y=L-0}=0,
\end{eqnarray}
see \cite{BMSV1} for details.

\section{Tensor modes and the hierarchy problem}
Let us study first the modes of the tensor field
$b_{\mu\nu}(x,y)$, which satisfies Eq. (\ref{ub}). Substituting
into this equation $$ b_{\mu\nu}(x,y) = c_{\mu\nu}e^{ipx}
\psi_n(y),\quad c_{\mu\nu} = const, \quad p^2 = -m_n^2, $$ we get:
\begin{eqnarray}\nonumber
\frac{d^2 \psi_n}{dy^2} -2(2(A')^2 -A'')\psi_n = -m_n^2 e^{2A}
\psi_n, \\ \label{bmode}  \psi_n^\prime |_{y=+0}= \psi_n^\prime +
2 A'  \psi_n |_{y=L-0}=0.
\end{eqnarray}
The boundary conditions suggest a substitution $\psi_n =
\exp(-2A)\omega_n $ (note that $A'|_{y=+0}=0$), which turns this
equation into
\begin{eqnarray}\nonumber
\frac{d}{dy} \left(e^{-4A}\omega_n'\right) = -m_n^2 e^{-2A}
\omega_n, \\ \label{bmode1}  \omega_n' |_{y=+0}=
\omega_n'|_{y=L-0}=0.
\end{eqnarray}
We see that the eigenfunctions $\omega_n$ are solutions of a
Sturm-Liouville problem with von Neumann boundary conditions. In
accordance with the general theory \cite{BKM}, the problem at hand
has no negative eigenvalues for arbitrary $A$, only one zero
eigenvalue, corresponding to $\omega_0 = const$.

The eigenfunctions $\{\psi_n(y)\}$ of eigenvalue problem
(\ref{bmode}) build a complete orthonormal set, the eigenfunction
of the zero mode being
\begin{equation} \label{zeromode}
\psi_0(y) = N e^{-2A(y)}.
\end{equation}
 Expanding  $b_{\mu\nu}$ in this system
\begin{equation}\label{decomp}
b_{\mu\nu}=\sum_{n=0}^\infty b_{\mu\nu}^n(x)\psi_n(y),
\end{equation}
we get four-dimensional tensor fields $b_{\mu\nu}^n(x)$ with
definite masses.

A standard technique gives us an expression for the
four-dimensional Planck mass on the brane
\begin{eqnarray}
M_{Pl}^{2}=M^{3}\int_{-L}^{L}e^{-2A}dy\simeq
M^{3}2e^{\frac{\left(\gamma
L\right)^{4}}{18M^{3}}}\int_{0}^{\infty}e^{-\frac{\left(\gamma
y\right)^{4}}{18M^{3}}}dy=\\
\nonumber =2M^{3}e^{\frac{\left(\gamma
L\right)^{4}}{18M^{3}}}\frac{\left(18M^{3}\right)^{\frac{1}{4}}}{4\gamma}\Gamma\left(\frac{1}{4}\right)\approx
3.7\cdot M^{3}e^{\frac{\left(\gamma
L\right)^{4}}{18M^{3}}}\frac{M^{\frac{3}{4}}}{\gamma}
\end{eqnarray}
and
\begin{eqnarray}
M_{Pl}\approx
2M\frac{M^{\frac{7}{8}}}{\sqrt{\gamma}}\,e^{\frac{\left(\gamma
L\right)^{4}}{36M^{3}}}.
\end{eqnarray}
Let us suppose that all fundamental parameters of the theory lie
in the $TeV$ range. To have the hierarchy problem solved, one
should take
\begin{eqnarray}
\frac{\gamma^{4}L^{4}}{36M^{3}}=\frac{\phi_{0}^{2}}{36M^{3}}\approx
36,
\end{eqnarray}
which means that
\begin{eqnarray}
\phi_{0}\simeq 36M^{\frac{3}{2}}
\end{eqnarray}
and
\begin{eqnarray}\label{L6}
L\simeq\frac{6M^{\frac{3}{4}}}{\gamma}.
\end{eqnarray}
Although Eq. (\ref{bmode1}) cannot be solved analytically for
$n\ne 0$, it is reasonable to suppose that the lowest masses of
the four-dimensional tensor excitations $b_{\mu\nu}^n(x)$ are of
the order of $L^{-1}$.

\section{Scalar sector and stability}
In order to find the mass spectrum of the scalar particles
described by Eq. (\ref{ug0}) let us substitute $$ g(x,y) = e^{ipx}
g_n(y), \quad p^2 = -\mu_n^2, $$ into this equation:
\begin{equation}\label{eq_sc}
g_n'' +2A'g_n'-2\frac{\phi''}{\phi'}g_n'-\frac{ (\phi')^2}{6M^3}
g_n= - \mu_n^2 e^{2A}g_n,
\end{equation}
\begin{eqnarray}
g'_n|_{y=+0}=0, \\ \label{bc11} \beta^{2}g'_n
-\mu_n^2e^{2A}g_n|_{y=L-0}=0.
\end{eqnarray}
It is necessary to note that since the field $f$ should be smooth
at the point $y=0$, from (\ref{gauge}) it follows that the value
$\left(g'_{n}/{\phi'}\right)'$ should be continuous at $y=0$ too.

First, let us solve Eq. (\ref{eq_sc}) for the case $\mu_{0}=0$,
i.e. for the zero mode. In the case of background solution
(\ref{bckgrsol}) the wave function $g_0$, satisfying boundary
condition at $y=0$, has the form
\begin{equation}
g_{0}\sim e^{-\frac{\gamma^{4}y^{4}}{18M^{3}}} +\frac{\gamma^{3}}
{\left(18M^{3}\right)^{\frac{3}{4}}}|y|^{3}
\int_{0}^{\frac{\gamma^{4}y^{4}}{18M^{3}}}q^{-\frac{3}{4}}e^{-q}dq.
\end{equation}
It is not difficult to check that $g'_{0}|_{y=L}\ne 0$. Thus, the
scalar zero mode is absent in the model.

Now let us examine, whether there are scalar tachyons in the
model. To this end we denote ${\tilde\mu}^{2}=-\mu^{2}>0$ (here
and below we omit the subscript $n$) and introduce a new
dimensionless variable $$t=\frac{\gamma}{M^{\frac{3}{4}}}y.$$ In
this case Eq. (\ref{eq_sc}) and boundary conditions take the form
\begin{eqnarray}\label{eq_sc1}
\ddot g + 2\dot
g\left(\frac{t^{3}}{9}-\frac{1}{t}\right)-\frac{2}{3}t^{2}g-{\bar\mu}^{2}\exp{\left(\frac{t^{4}}{18}\right)}g=0,
\\ \label{gcondt}
\left. \frac{M^{\frac{3}{4}}\beta^{2}}{\gamma}\dot
g+{\bar\mu}^{2}\exp{\left(\frac{t^{4}}{18}\right)}g\right|_{t=\frac{\gamma
L}{M^{\frac{3}{4}}}}=0,\\ \dot g|_{t=0}=0,
\end{eqnarray}
where
$\bar\mu=\tilde\mu\frac{M^{\frac{3}{4}}}{\gamma}\exp{\left(-\frac{\left(\gamma
L\right)^{4}}{36M^{3}}\right)}$, $\frac{\gamma
L}{M^{\frac{3}{4}}}\approx 6$ (see (\ref{L6})) and $\dot{g}\equiv
{dg}/{dt}$.

\begin{figure}[pb]
\centering
\includegraphics[width=16cm]{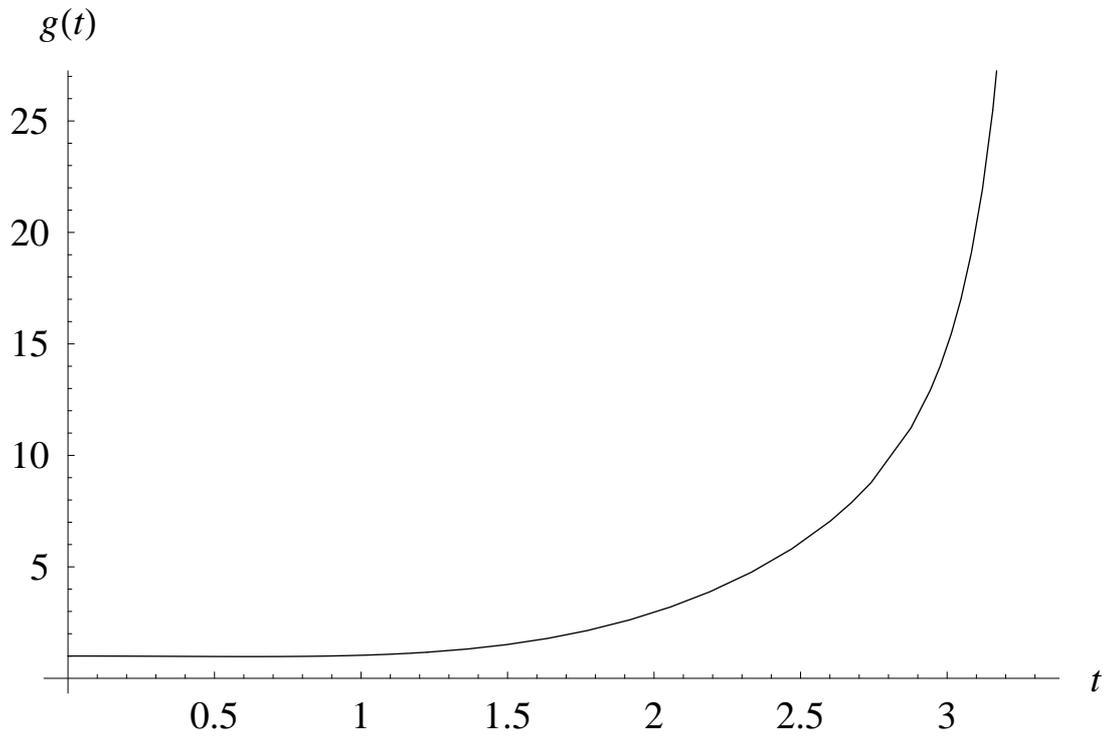}
\caption{Numerical solution for $g(t)$, $\bar\mu=0.5$}
\label{fig1}
\end{figure}

\begin{figure}[pb]
\centering
\includegraphics[width=16cm]{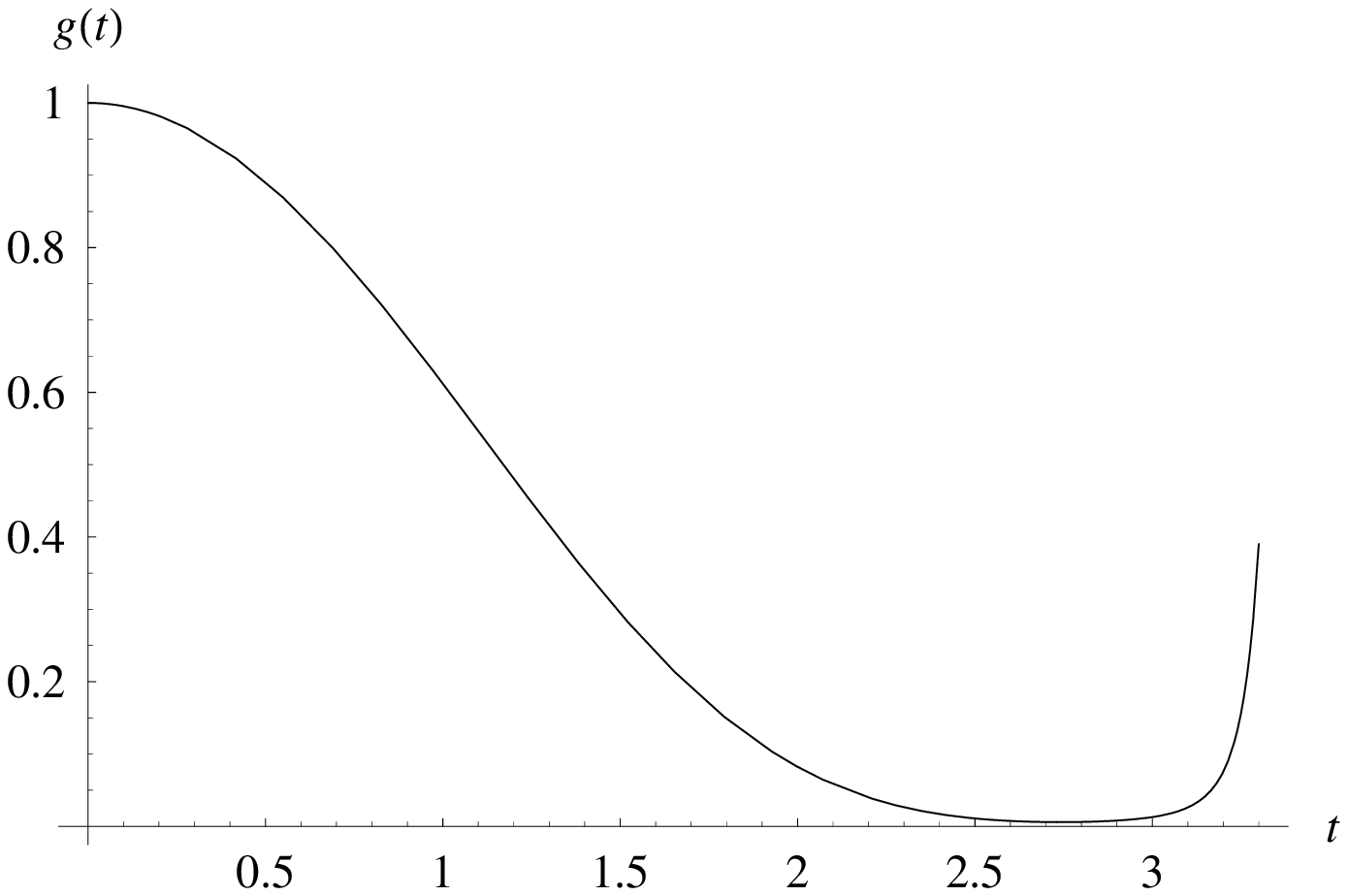}
\caption{Numerical solution for $g(t)$, $\bar\mu=0.9507$}
\label{fig2}
\end{figure}

\begin{figure}[pb]
\centering
\includegraphics[width=16cm]{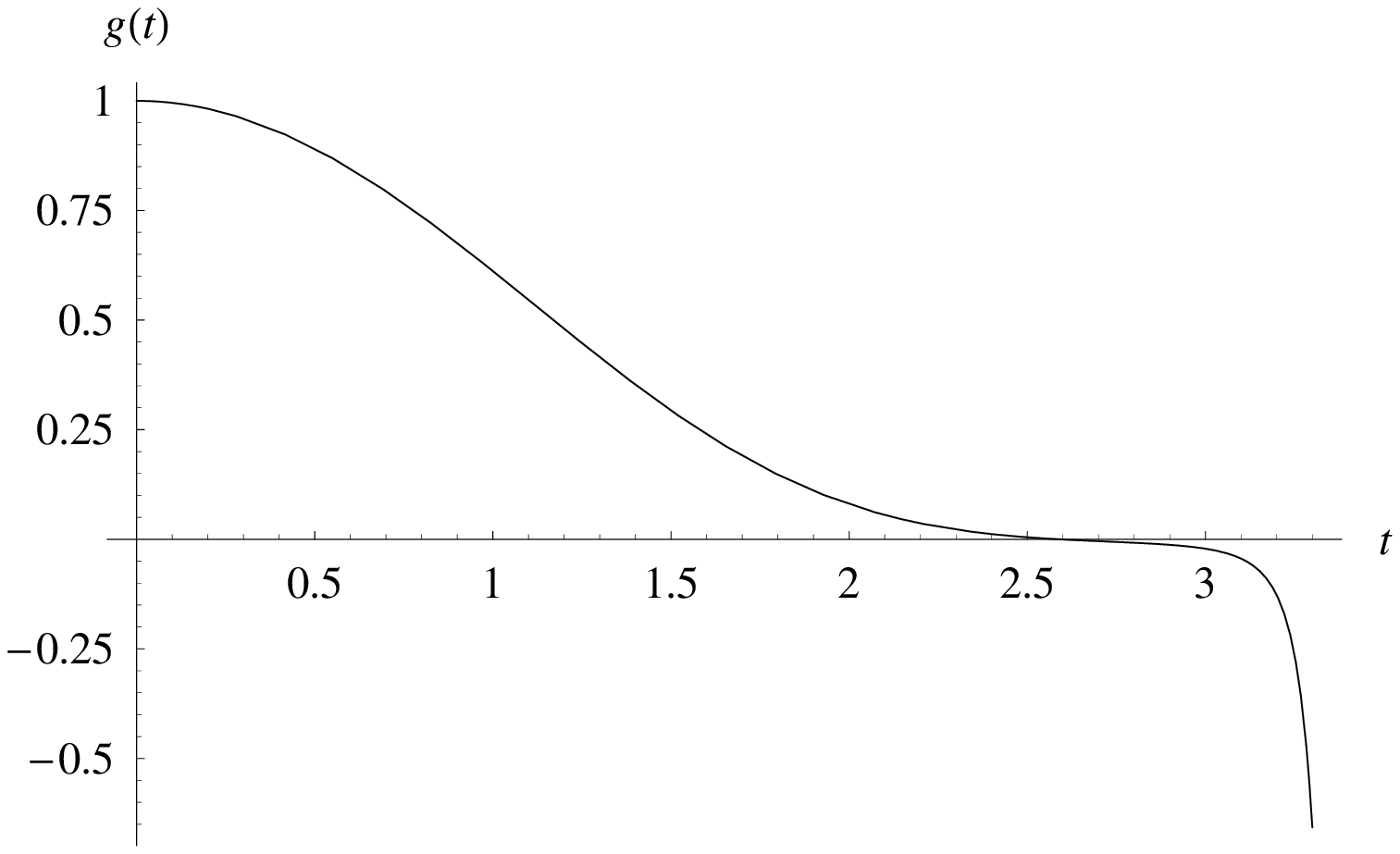}
\caption{Numerical solution for $g(t)$, $\bar\mu=0.9509$}
\label{fig3}
\end{figure}

\begin{figure}[pb]
\centering
\includegraphics[width=16cm]{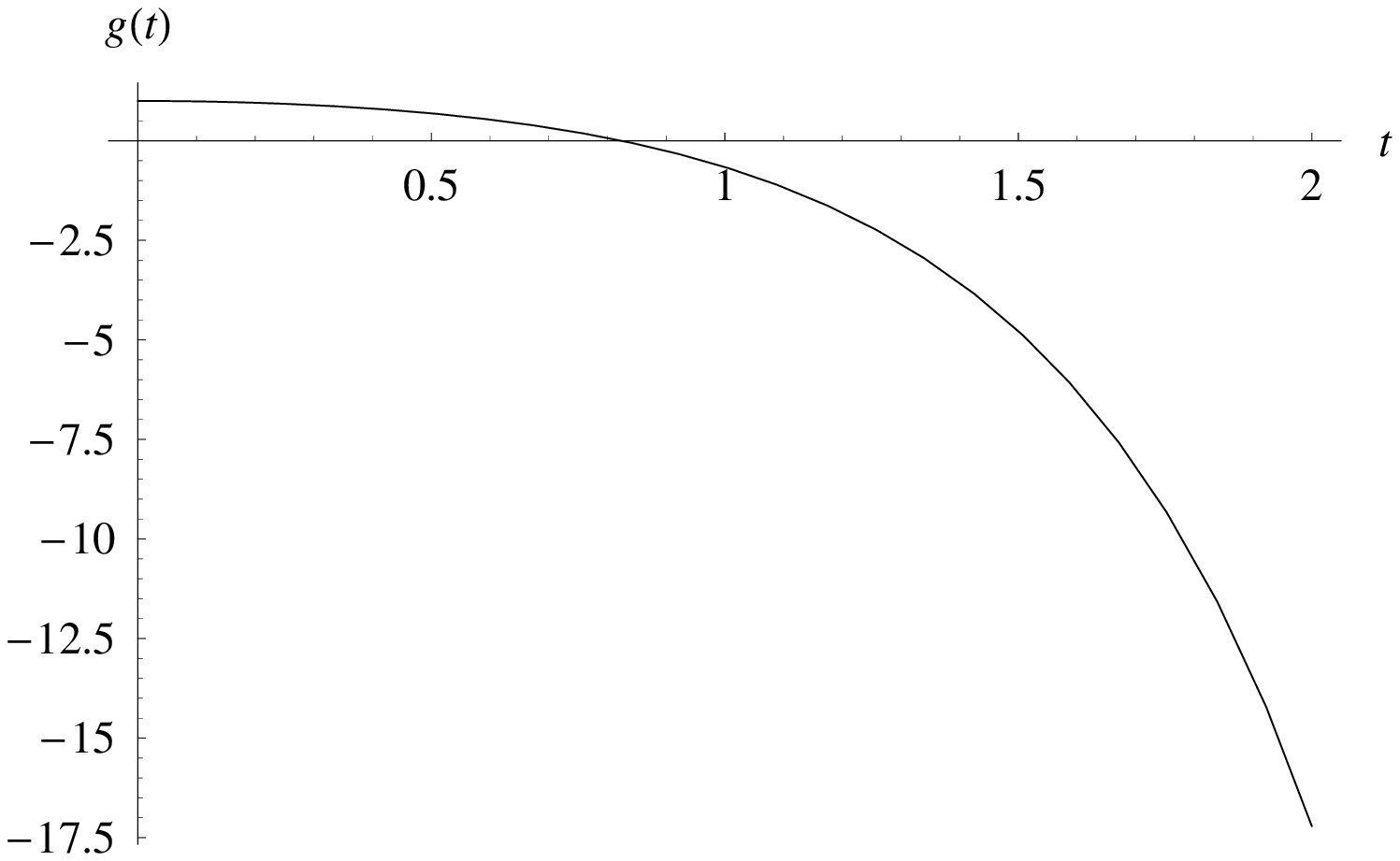}
\caption{Numerical solution for $g(t)$, $\bar\mu=1.5$}
\label{fig4}
\end{figure}

Unfortunately we cannot solve Eq. (\ref{eq_sc1}) analytically.
Numerical analysis (see Appendix A) shows that for $\bar\mu\le
0.9507$, $t=3.2$: $g(t)>0$ and $\dot g(t)>0$ (see examples on
Figs. \ref{fig1}, \ref{fig2}). At the same time for $\bar\mu\ge
0.9509$, $t=3.2$: $g(t)<0$ and $\dot g(t)<0$ (see examples on
Figs. \ref{fig3}, \ref{fig4}; Fig. \ref{fig4} is shown for $t\leq
2$, but one can check that for $\bar\mu=1.5$ and $t=3.2$ \,
$g(t)<0$ and $\dot g(t)<0$). The graphs on Figs. \ref{fig1},
\ref{fig2}, \ref{fig3} are shown for $t\le 3.3$, it is made to
show the behavior of $g(t)$ in the interval $t\in [0,3.2]$,
especially for the cases $\bar\mu=0.9507$ and $\bar\mu=0.9509$.
For $t>3.2$: $\dot g/g>0$, it can be easily seen from the
structure of Eq. (\ref{eq_sc1}). Indeed, let us divide
(\ref{eq_sc1}) by $g$ and pass to the equation for $q(t)=\dot
g/g$, which takes the form of a Riccati equation
\begin{equation}\label{eqforq}
\dot q +
q^{2}+2\left(\frac{t^{3}}{9}-\frac{1}{t}\right)q=\frac{2}{3}t^{2}+{\bar\mu}^{2}\exp{\left(\frac{t^{4}}{18}\right)}.
\end{equation}
If initially for some value of $t=t_{q}>0$: $q>0$, then $q$ will
remain positive for any $t>t_{q}$. Indeed, the function $q$ should
pass through zero to change the sign. But if $0<q\ll 1$, then from
Eq. (\ref{eqforq}) it follows that $\dot q>0$, $q$ appears to be a
growing function and thus remains positive. Thus, for $t>3.2$:
$\dot g(t)>0$, $g(t)>0$ or $\dot g(t)<0$, $g(t)<0$ depending on
the sign of $g(t)$ at $t=3.2$. In both cases (\ref{gcondt}) is not
satisfied, since $M>0$, $\gamma>0$ and $\beta^{2}>0$.

Eq. (\ref{gcondt}) can be satisfied if $\dot g(t)=g(t)=0$ at the
point $t=\frac{\gamma L}{M^{\frac{3}{4}}}\approx 6$. But such
boundary conditions imply that $g(t)\equiv 0$ in the interval
$t\in \left[\epsilon, \frac{\gamma L}{M^{\frac{3}{4}}}\right]$ for
any $\epsilon>0$ -- this conclusion follows from the theorem about
the existence and uniqueness of solution for the Cauchy problem --
see, for example, \cite{elsgolts}. Finally, due to the
continuity of the function $g(t)$, we get $g(t)\equiv 0$ in the
interval $t\in \left[0,\frac{\gamma L}{M^{\frac{3}{4}}}\right]$.
Of course, the same conclusion can be made for any
$0<t_{1}<\frac{\gamma L}{M^{\frac{3}{4}}}$ such that $\dot
g(t_{1})=g(t_{1})=0$.

As for the region $0.9507<\bar\mu<0.9509$, we have made a large
number of numerical simulations with different values of
$\bar\mu$. The behavior of the corresponding solutions is such
that for $t\ge 3.6$: $g(t)>0$, $g(t)>0$ or $\dot g(t)<0$, $g(t)<0$
respectively, analogous to the behavior of solutions presented on
Figs. \ref{fig2}, \ref{fig3}. A simple qualitative explanation of
this fact can be given. For $0.9507<\bar\mu<0.9509$ and $t>3.6$
the coefficient ${\bar\mu}^{2}\exp{\left(\frac{t^{4}}{18}\right)}$
in (\ref{eq_sc1}) grows rapidly, which leads to the growth of the
absolute value of function $g(t)$ with coordinate $t$ for $t>3.6$.

Nevertheless, for some value of $\bar\mu$ such that
$0.9507<\bar\mu<0.9509$ there exists a solution, which satisfies
condition (\ref{gcondt}). Indeed, let us define a function
$F(\bar\mu)=\frac{M^{\frac{3}{4}}\beta^{2}}{\gamma}\dot
g(t)+{\bar\mu}^{2}\exp{\left(\frac{t^{4}}{18}\right)}g(t)|_{t=\frac{\gamma
L}{M^{\frac{3}{4}}}}$. For $\bar\mu<0.9507$ it is positive,
whereas for $0.9509<\bar\mu$ it is negative (see Figs. \ref{fig1},
\ref{fig2}, \ref{fig3}, \ref{fig4}). Thus, it is reasonable to
suppose that there exists an appropriate value $\bar\mu^{*}$
($0.9507<\bar\mu^{*}<0.9509$) such that
$F(\bar\mu^{*})|_{t=\frac{\gamma L}{M^{\frac{3}{4}}}}=0$, which
corresponds to a tachyonic mode. It appears to be very difficult
to find the exact value of the tachyonic mass numerically. At the
same time the physical mass of the tachyon is such that
\begin{equation}
{\mu^{*}}^{2}\approx
-{\bar\mu}^{*2}M_{Pl}^{2}\frac{\gamma^{3}}{3.7\cdot
M^{\frac{21}{4}}}\approx -\frac{0.9}{3.7} M_{Pl}^{2}\approx
-\left(0.5\cdot 10^{19}GeV\right)^{2}
\end{equation}
for the  given values of the model parameters (we suppose that
$\gamma\approx M^{\frac{7}{4}}$). Such energy scale lies outside
the range of validity of our effective theory, described by action
(\ref{actionDW}) (because $\left|\mu^{*}\right|\sim E\gg M$, see
Section~2). From the classical point of view it can be understood
as follows: the tachyonic mode should behave as $e^{\mu^{*}
x^{0}}$. The time derivative of the tachyon field $\sim\mu^{*}
e^{\mu^{*} x^{0}}$, i.e. it is enhanced by the large value of
$\mu^{*}\sim M_{Pl}$ in comparison with the tachyon field itself.
Thus, even if the value of the tachyon field is small, its time
derivative would lead to breakdown of perturbative approach and
corresponding nonlinear effects, coming from the five-dimensional
curvature (through substitution (\ref{subst})). Another remarkable
thing is that the wave function of the tachyon is such that if
$g|_{y=0}=1$, then in the leading order
$g|_{y=L}\sim\exp\left(-\mu^{*} \exp\left({\frac{(\gamma
L)^{4}}{36M^{3}}}\right)\right)\approx \exp\left(-\mu^{*}
\exp\left(36\right)\right)$. It means that when the nonlinear
effects and (or) effects of the underlaying fundamental theory
begin to  affect the behavior of the theory in the bulk, the
theory on the brane remain intact, because the coupling constant
of the tachyon to matter on the brane, which is proportional to
the value of the wave function on the brane, is negligibly small -
much smaller than the coupling constant of the massless tensor
graviton. Thus, the runaway of the scalar field can be stopped in
the bulk because of the nonlinear effects coming from  action
(\ref{actionDW}) or from the underlaying fundamental theory. Of
course, we cannot argue that it is indeed so, but such situation
can be realized.

Of course, our examination is not explicit since it is based on
the numerical calculations. But we think that the analysis made
testifies in favor of absence of scalar tachyons in the model
below the energy scale of our effective theory (\ref{actionDW}).
As for the ghosts, the form of the effective action for the scalar
modes ensures the proper signs of the appropriate kinetic
terms \cite{BMSV1}.

The form of Eq. (\ref{eq_sc1}) allows us to estimate the mass of
the lowest scalar excitation and its coupling to matter on the
brane. Indeed, let us suppose that the lowest mass $\mu_{1}$ (see
(\ref{eq_sc})) is such that $\mu_{1}/M\approx O(1)$. In this case
we can neglect the last term in Eq. (\ref{eq_sc}) in comparison
with the last but one term of this equation, and the solution of
the resulting equation takes the form
\begin{equation}\label{zero-one-mode}
g_{1}(y)\approx A_{1}\left( e^{-\frac{\gamma^{4}y^{4}}{18M^{3}}}
+\frac{\gamma^{3}} {\left(18M^{3}\right)^{\frac{3}{4}}}|y|^{3}
\int_{0}^{\frac{\gamma^{4}y^{4}}{18M^{3}}}q^{-\frac{3}{4}}e^{-q}dq\right)\sim
g_{0}(y),
\end{equation}
where $A_{1}$ is a normalization constant. Let us suppose that the
size of the extra dimension is such that $\gamma
L/M^{\frac{3}{4}}=6$, see (\ref{L6}). The values $g_{1}(L)$ and
$g_{1}'(L)$ can be easily calculated, which gives us
\begin{eqnarray}\label{estim1}
g_{1}(L)\approx A_{1}\cdot 89.6,\\ \label{estim2} g_{1}'(L)\approx
A_{1}\cdot 44.8\frac{\gamma}{M^{\frac{3}{4}}}.
\end{eqnarray}
Substituting (\ref{estim1}) and (\ref{estim2}) into (\ref{bc11})
we easily get
\begin{equation}
\mu_{1}^{2}\simeq\frac{\beta^{2}\gamma}{2M^{\frac{3}{4}}}.
\end{equation}
For example, if $\beta^{2}\simeq M$, $\gamma\simeq M^{7/4}$ and
$M\approx 10\, TeV$, the lowest mass $\mu_{1}\approx 7\, TeV$. Of
course, it can be even smaller depending on the values of the
parameters $\beta$, $\gamma$ and $M$.

It is also necessary to note that the analysis carried out with
the help of the numerical solution of Eq. (\ref{eq_sc}) for such
small $\mu_{1}$ reproduces the results obtained using
(\ref{zero-one-mode}) with a very good accuracy (of the order of
$1-2\%$).

Now let us calculate the coupling constant of the first scalar
mode to matter on the brane. To this end we need to calculate the
normalization constant $A_{1}$. The normalization condition for
the scalar modes takes the form \cite{BMSV1}
\begin{equation}
\int_{0}^{L}
dye^{2A}\left(g_{1}^{2}+\frac{6M^{3}}{(\phi')^2}{g'_{1}}^{2}\right)=\frac{2}{3}.
\end{equation}
It is more convenient to pass to the variable
$t=\frac{\gamma}{M^{\frac{3}{4}}}y$:
\begin{equation}\label{int-t}
e^{-72}\int_{0}^{6} dt
e^{\frac{t^{4}}{18}}\left(g_{1}^{2}+\frac{6}{4t^{2}}\dot
g_{1}^{2}\right)=\frac{2\gamma}{3M^{\frac{3}{4}}}.
\end{equation}
The integral in (\ref{int-t}) can be evaluated numerically, which
gives us
\begin{equation}
A_{1}^{2}\approx 0.004\frac{\gamma}{M^{\frac{3}{4}}}.
\end{equation}
Now we can calculate the coupling constant of the lightest scalar
mode to matter on the brane (see \cite{BMSV1}):
\begin{equation}
\epsilon_{1}=-\frac{g_{1}(L)}{2\sqrt{8M^{3}}}\approx
-\frac{A_{1}\cdot 89.6}{2\sqrt{8M^{3}}}\approx
-\sqrt{\frac{\gamma}{M^{\frac{15}{4}}}}.
\end{equation}
One can see that $\epsilon_{1}\approx \frac{-1}{10\, TeV}$ for the
given values of the fundamental parameters $\gamma$ and $M$.

Unfortunately it is impossible to calculate even the lowest mass
of the tensor excitations using the method described above. One
should carry out a very precise numerical analysis to get an
information about the spectrum of the tensor modes.

\section{Conclusion}
In this paper a model describing the scalar field minimally
coupled to gravity in the spacetime with one compact extra
dimension is proposed. It admits the existence of a single
tensionful brane, contrary to the most brane world models with one
compact extra dimension demanding the existence of at least two
branes (of course, except the simplest case of the ADD
model \cite{ADD,Ant} with tensionless branes). We also showed that
the model could be interesting in  view of the hierarchy problem.

The linearized gravity in the model was studied under the
assumption of the  $Z_{2}$ orbifold symmetry. We obtained the
expression for the four-dimensional Planck mass on the brane in
terms of the fundamental five-dimensional parameters of the
theory. We also made a stability analysis of the model, --
analytical for the tensor modes and numerical for the scalar
modes, which resulted in the conclusion that the scalar sector of
the model contain one tachyon, which corresponds to the result
obtained in \cite{Lesgourgues:2003mi}, its "mass" being of
the order of the four-dimensional Planck mass. Thus, the model as
it is, at least in the linear approximation, is unstable and its
"lifetime"\ is of the order of the four-dimensional Planck time.
Nevertheless, the energy scale of the tachyon is such that
multidimensional nonlinear fundamental underlaying theory can come
to play and "lift up" the scalar sector from falling down. Of
course it is not necessarily so, but in principle it seems to be
possible.

The background solution can also be used to describe the world
with two branes. Indeed, the second brane can be placed at the
point $y=L_{0}$, $0<L_{0}<L$. The results of \cite{BMSV1}
suggest that such a system is totally devoid of tachyons.

It is very interesting to carry out a numerical calculation of the
coupling constants and the masses of the tensor modes and a
complete description of the scalar sector of the model, as well as
the model without $Z_{2}$ orbifold symmetry (in the latter case
there should appear antisymmetric modes). One can also use the
model discussed in this paper (for example, the stable
configuration with two branes) as a basis for constructing models
with universal extra dimensions. These tasks deserve additional
thorough investigation.

\section*{Acknowledgments}

The work was supported by grant of Russian Ministry of Education
and Science NS-8122.2006.2. M.S. also acknowledges support of
grant for young scientists MK-8718.2006.2 of the President of
Russian Federation, grant of the "Dynasty" Foundation and
scholarship for young teachers and scientists of M.V.~Lomonosov
Moscow State University. The authors are grateful to R.~Bogdanov,
D.~Levkov, M.~Libanov, V.~Rubakov and V.~Shakhparonov for valuable
discussions.

\section*{Appendix: note on numerical analysis}
We solve Eq. (\ref{eq_sc1}) numerically with the following initial
conditions on the "time"\ variable $t$:
\begin{eqnarray}
g(t)|_{t=0}=1,\\
\dot g(t)|_{t=0}=0.
\end{eqnarray}
But since the coefficient
$$\frac{t^{3}}{9}-\frac{1}{t}$$ in (\ref{eq_sc1})
is not defined at the point $t=0$, it is inconvenient to use the
point $t=0$ as the initial point for numerical calculations. To
bypass this problem, we find an approximate analytical solution of
Eq. (\ref{eq_sc1}) in the vicinity of the point $t=0$:
\begin{eqnarray}
g(t)\approx 1-\frac{\bar\mu^{2}}{2}t^{2}.
\end{eqnarray}
Now we choose the point $t_{0}=10^{-11}$ as the initial point
instead of $t=0$. The corresponding initial conditions take the
form
\begin{eqnarray}
g(t)|_{t=t_{0}}=1-\frac{\bar\mu^{2}}{2}10^{-22},\\
\dot g(t)|_{t=t_{0}}=-\bar\mu^{2}10^{-11}.
\end{eqnarray}
The numerical analysis for a large number of different values of
$\bar\mu$ was made using the program package {\it Mathematica},
version 5.2. Selected solutions are presented on Figs. \ref{fig1},
\ref{fig2}, \ref{fig3}, \ref{fig4}.

\end{document}